# On the Time-Range Dependency of the Beampatterns Produced by Arbitrary Antenna Arrays: Discussions on the Misplaced Expectations from Frequency Diverse Arrays

M. Fartookzadeh

Abstract: In the recent years constructing time-invariant spatial-focusing beampatterns by using frequency diverse arrays (FDAs) is conveyed in several papers. Further than time-invariant spatial-focusing beampatterns, the unfeasibility of constructing range-dependent time-invariant beampatterns is demonstrated in this paper. It is indicated that the beampatterns produced by antenna arrays including FDAs are not independently controllable in time, $t$, and range, $r$, dimensions, in the farfield. Inappropriate applications of the FDAs having time-invariant spatial-focusing beampatterns in two recent papers are also discussed.

## I. Introduction

Frequency diverse arrays (FDAs) with time-invariant spatial-focusing beampatterns could be of high importance for data transfer and radar applications. However, a time-invariant spatial-focusing beampattern can never be constructed in the farfield due to the fundamental issues in physics. Yet, constructing the time-invariant spatial-focusing beampatterns by using FDAs has been studied, extensively in the recent years. For example, application of the time-invariant focusing FDA and polarization-subarray based radar for improving signal to interference ratio is studied in [1]. Other applications of FDAs with time-invariant spatial-focusing patterns could be in secure wireless information transfer and remote power transmission [2, 3].

The main reason of incorrectly obtaining time-invariant spatial-focusing beampatterns in the recent works including [1-5] is an incorrect definition of time in some equations. In fact, the delay of traveling electromagnetic wave, $r/c$, to the distance $r$ from the array point is neglected in calculating the transmitted signal ($c$ is the speed of electromagnetic wave). The detailed explanations on the previous works with the similar incorrect definition of time for obtaining time-invariant spatial-focusing beampatterns and why obtaining this kind of beampatterns is impossible are available in [6-8].

In this paper, a general proof is proposed to indicate that further than the spatial-focusing beampatterns; constructing range-dependent time-invariant beampatterns is not possible in the farfield. The proof is presented in the following section, and the incorrect parts of [1, 3] are specified and discussed in section III.

## II. Range-Time Dependency of Arrays

A general proof of the unfeasibility of constructing range-dependent time-invariant beampatterns is provided in this section based on the beampattern of a single antenna element. The range dependency of the received signal in the farfield range $R_0$, from an antenna excited by the signal $S_i^0(t)$ is

$$S_r^0(t, R_0) = \frac{k}{(R_0)^2} S_i^0\left(t - \frac{R_0}{c}\right), \tag{1}$$

M. Fartookzadeh is with Department of Electrical and Electronics Engineering, Malek Ashtar University, P. O. Box 1774-15875, Tehran, Iran (Mahdi.fartookzadeh@gmail.com).

where $k$ is a constant depending on the characteristics of the antenna elements and the receiver, and $1/(R_0)^2$ is emitted in solving the wave equations in spherical coordinates, the well-known inverse-square law corresponding to free-space path loss (FSPL). The range-time dependency of the beampattern is therefore only the $R_0/c$ delay from the initial signal at the array point, since the $k/(R_0)^2$ factor is not considered in calculating the beampattern. Admitting this kind of range-time dependency for a single antenna the array case will be discussed as following.

An N-element linear array excited by arbitrary signals is considered to indicate the time-range dependency of all kind of antenna arrays including FDAs. Schematic of the array with separation $d$ between elements is indicated in Fig. 1. The difference between delay of each element and a point in farfield with the next element is $d\sin\theta/c$, which can be compensated in the input signals of each element to make a unique delay between the signal sources and the observation point. The received signal to a point in farfield at the distance $R_0$ from the array is therefore obtained from

$$\begin{aligned} S_r(t_1, R_0) &= \sum_{n=0}^{N-1} \frac{k_n}{(R_n)^2} S_i^n\left(t_1 - \frac{R_n}{c}\right) \\ &= \sum_{n=1}^{N} \frac{k_n}{(R_0 - nd\sin\theta)^2} S_i^n\left(t_1 - \frac{R_0 - nd\sin\theta}{c}\right). \end{aligned} \tag{2}$$

where $k_n$ can be divided into the transmitter part $k_t^n$, and the receiver part $k_r$, which is independent of $n$. Now by considering the farfield condition, $R_0 \gg (N-1)d$, and assuming compensated input signals as

$$S''^n_i(t_1) = S'^n_i\left(t_1 + \frac{nd\sin\theta}{c}\right) = k_t^n S_i^n\left(t_1 + \frac{nd\sin\theta}{c}\right), \tag{3}$$

the received signal will be

$$S_r(t_1, R_0) = \frac{k_r}{(R_0)^2} \sum_{n=0}^{N-1} S''^n_i\left(t_1 - \frac{R_0}{c}\right). \tag{4}$$

The range dependency of the received signal is only the inverse-square $1/(R_0)^2$, and the $R_0/c$ delay from the initial signal at the array point.

Consequently, if two points at different ranges $R_1$ and $R_2$ are assumed with similar angles, $\theta$, the received signal to $R_2$ at $t = t_1$ is exactly same as the received signal to $R_1$ at $t = t_1 - \Delta R/c$ multiplied by $(R_1/R_2)^2$,

$$S_r(t_1,\ R_2) = \left(\frac{R_1}{R_2}\right)^2 S_r\left(t_1 - \frac{\Delta R}{c},\ R_1\right), \tag{5}$$

and the signal received to $R_1$ at $t_1$ will be received on $R_2$ at $t = t_1 + \Delta R/c$ as indicated in Fig. 2. Therefore, the beampatterns produced by antenna arrays including FDAs are not independently controllable in time, $t$, and range, $r$, dimensions, in the farfield. The beampatterns are indeed functions of $t - r/c$, and hence constructing range-dependent time-invariant beampatterns in the farfield is impossible.

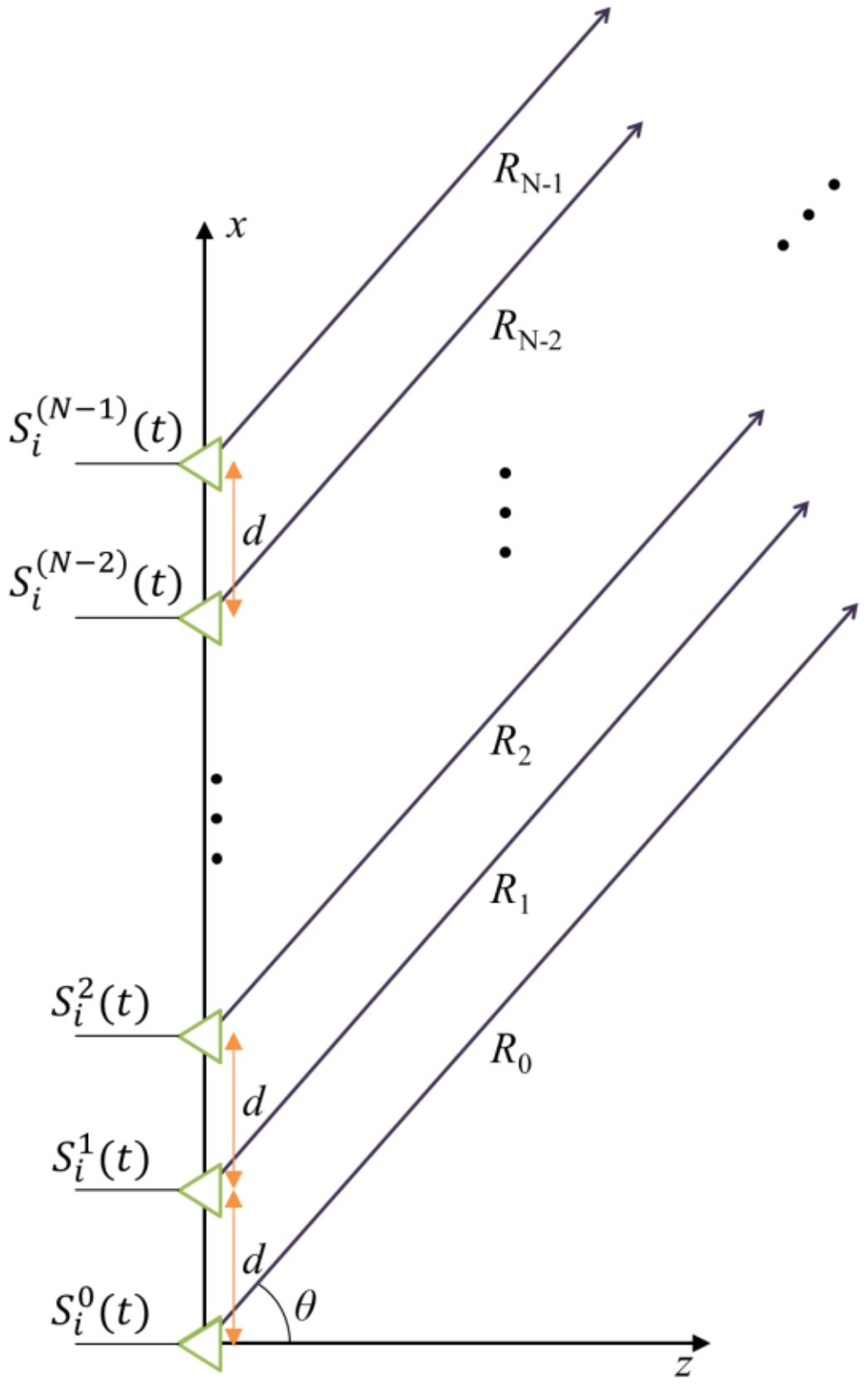

Fig. 1. Schematic of the N-element linear array excited by arbitrary signals.

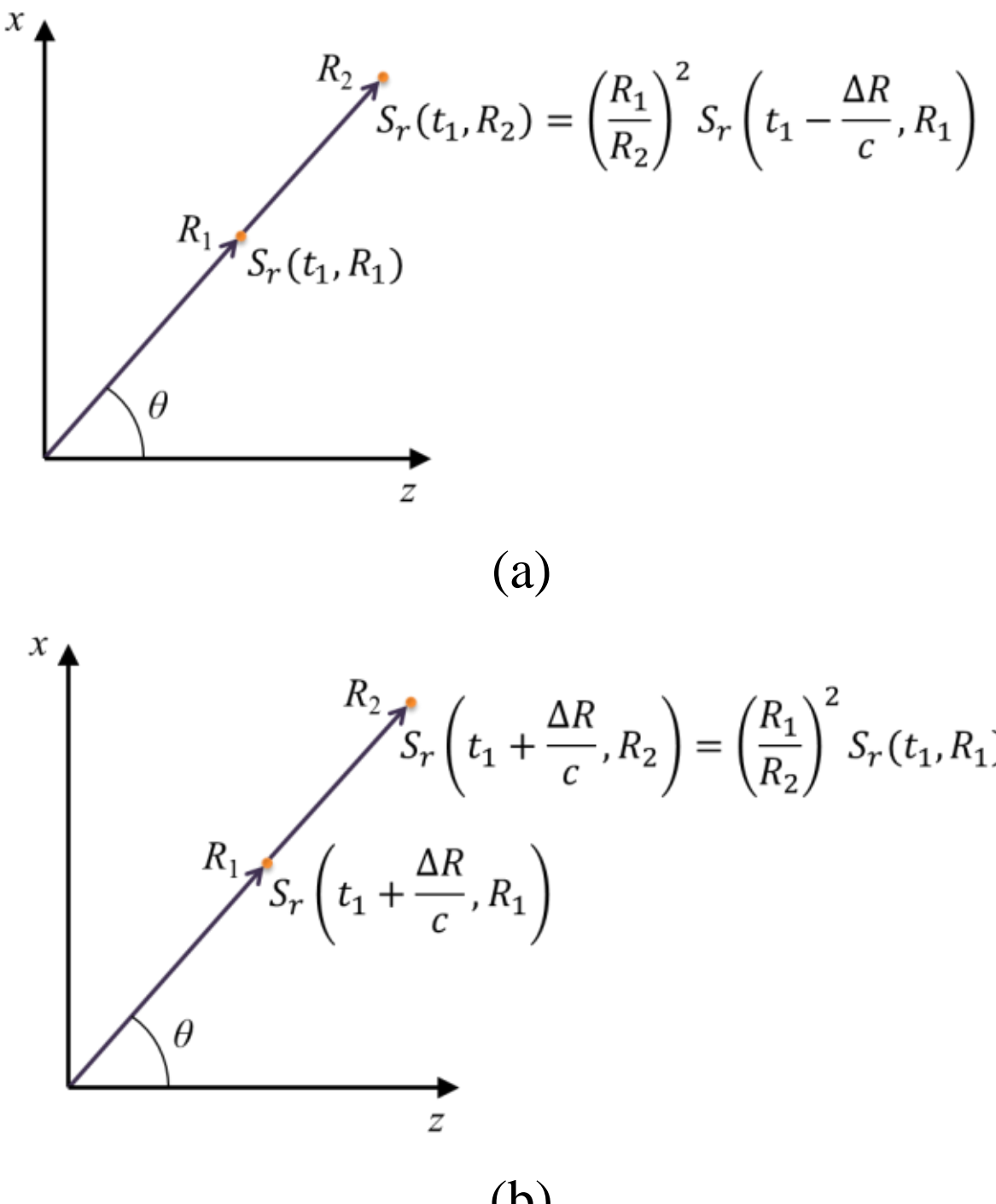

Fig. 2. The received signals to the two points at two different ranges $R_1$ and $R_2$ with similar angles, $\theta$, at (a) $t = t_1$ and (b) $t = t_1 + \Delta R/c$ ($\Delta R = R_2 - R_1$).

## III. Inappropriate Applications of FDAs

Inappropriate applications of the FDAs with time-invariant spatial-focusing beampatterns in two recent papers are discussed in this section. The first application is the improvement of signal to interference ratio in [1]. The main source of obtaining incorrect results in [1] is the section II-B, which introduces the time-modulated frequency-offsets. In particular, in Eq. (28) of [1], $\Delta f_i(t), i = 1,2$, should be replaced by $\Delta f_i(t - R/c)$, leading to the movement of focusing point with time. Producing angle-range-focused (dot-shaped) beampatterns with FDAs in an instance of time is not in contrast with the explanations in section II, while the time dependency is the main conflict. Consequently, comparison between the spatial beampatterns of the target and the interference in section IV of [1] is also incorrect, since the beampatterns at the target point are obtained by incorrect assumption of the time.

Simultaneous wireless information and power transfer (WIPT) by using retrodirective FDA is the next infeasible application of FDAs with time-invariant spatial-focusing beampatterns. The idea is to add tiny frequency offsets across the phase-conjugating mixers in the retrodirective array to produce time-invariant range-angle focused beampatterns [3]. It could be obtained if the beampatterns produced by FDAs are independently controllable in time, $t$, and range, $r$, dimensions. However, the beampatterns (BP) produced by FDAs are not functions of $t$ and $r$ independently, $BP(t, r)$, and they are functions of $t - r/c$, $BP(t - r/c)$, causing the movement of beampatterns with time. In addition, constructing range-invariant time-invariant beampatterns is feasible by using traditional arrays and does not require FDAs. Furthermore, the time-invariant title for the beampattern in Fig. 10 of [3] is evidently incorrect.

For 'instantaneous time' patterns, such as the patterns in Fig. 3, Fig. 9 and Fig. 10 (a) of [3], the time at which the patterns are plotted should be determined, since the patterns are not constant and moving with time. In addition, Fig. 10 (b) of [3], should be change to something like Fig. 3. And we can say the 'instantaneous time' pattern in Fig. 10 (a) of [3] is obtained at $t = 0.7 \times 10^{-5}$ s (if we assume the excitation begins at $t = 0$). Also, the 'time-variant' pattern in Fig. 10 is for $t \in [0.65, 0.75] \times 10^{-5}$ s.

Unfortunately, some parts of the extensive works in [1-5] are based on several published papers having incorrect assumption of the time in calculating the FDA beampatterns [6-8].

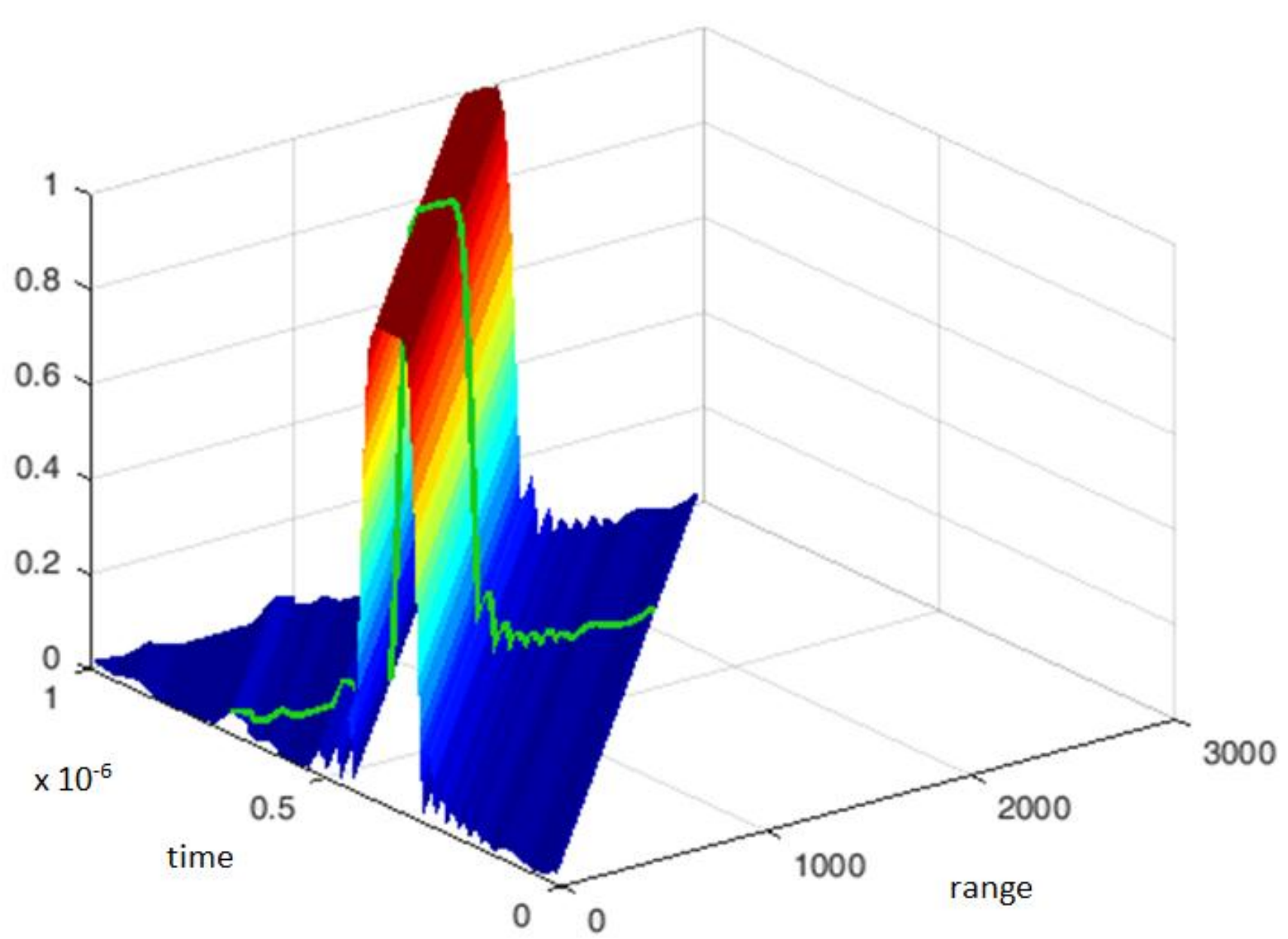

Fig. 3. Time-variant pattern corresponding to Fig. 10 (b) of [3]

# IV. Conclusions

The unfeasibility of constructing range-dependent time-invariant beampatterns in the farfield has been proven in this paper. In addition, explanations have been presented on the inappropriate applications of the FDAs in two recent papers concerning time-invariant spatial-focusing beampatterns. As a consequence, producing dot-shaped beampatterns with FDAs are possible in an instance of time, while it is not possible to construct time-invariant focused beampatterns in farfield. Furthermore, constructing a range-dependent time-invariant beampattern is impossible, since the beampattern in time behave exactly same as in range divided by $c$.